\documentclass[]{aa}

\usepackage{graphicx}
\usepackage{natbib}
\usepackage{amssymb}
\usepackage{caption}

\begin{document}

\title{Optical flux behaviour of a sample of Fermi blazars}

 \author{E. J. Marchesini\inst{1,2}
   \and I. Andruchow\inst{1,2}%
   \and S. A. Cellone\inst{1,2}
   \and J. A. Combi\inst{1,3}
   \and L. Zibecchi\inst{1,2}
   \and J. Mart\'{\i}\inst{4,6}
   \and G. E. Romero\inst{1,3}
   \and A. J. Mu\~noz-Arjonilla\inst{6}
   \and P. Luque-Escamilla\inst{5,6}
   \and J. R. S\'anchez-Sutil\inst{6}}
   \institute{Facultad de Ciencias Astron\'omicas y Geof\'{\i}sicas,
     Universidad Nacional de La Plata, Paseo del Bosque, B1900FWA La Plata,
     Argentina.
\and IALP, CONICET-UNLP, CCT La Plata, Paseo del Bosque, B1900FWA - La Plata -
Argentina.
     \and IAR, CONICET, CCT La Plata, C.C. No. 5 (1894) Villa Elisa - Buenos
     Aires - Argentina.
\and Departamento de F\'{\i}sica (EPS), Universidad de Ja\'en, Campus Las
Lagunillas s/n, A3, 23071 - Ja\'en - Spain.
\and Dept. de Ingenier\'{\i}a Mec\'anica y Minera, EPS de Ja\'en,  Universidad de Ja\'en, Campus Las Lagunillas s/n, A3-402, 23071 Ja\'en, Spain.
\and Grupo de Investigaci\'on FQM-322, Universidad de Ja\'en, Campus Las Lagunillas s/n, A3-065, 23071 Ja\'en, Spain.
}

\abstract{}{We aim at investigating the time-behaviour of a sample of gamma-ray blazars. We present the results from a 13 month-long optical photometry
  monitoring campaign of the blazars PKS\,0048$-$097, PKS\,0754$+$100,
  $[$HB89$]$\,0827$+$243, PKS\,0851$+$202, PKS\,1253$-$055, PKS\,1510$-$089,
  PKS\,1749$+$096, PKS\,2230$+$114 and PKS\,2251$+$158.}{We
  analyse the variability of each object, focusing on different time-scales (long term, short term, and microvariability), in an attempt to
  achieve a statistical comparison of the results.} {After applying a
  geometric model to explain the variability results, we found that it is
  possible that a slight change in the direction of the jet generates the variations detected in some objects during this campaign.}  {}


\maketitle

\section{Introduction}

Blazars are a sub-type of active galactic nuclei (AGN), whose jets axes are extremely close to the
line of sight; the electromagnetic (non-thermal) emission from the jet is
thus relativistically boosted and dominates along the whole spectrum. These
objects often show a flat radio spectrum and apparent superluminal motions
at VLBI scales \citep{Urry99}, while both short- and long-term variations are
detected in optical flux \citep[e.g.][]{Miller89, Carini90, Carini92,
  Romero02}. Most blazars present microvariability, e.g.: variations over
intranight timescales. These changes in brightness can also be detected at
other wavelengths, from radio to gamma rays.

Blazars are subclassified as flat spectrum radio quasars (FSRQ) and BL\,Lac
objects (BL\,Lacs). In the first case, strong emission lines are present in
their optical spectra while, in the second, these lines are weak or
non-existent. A (possibly) more physically based classification takes into
account the position of the synchrotron peak (SP) in the spectral energy
distribution (SED): in this way, blazars are classified as high- (HSP),
intermediate- (ISP) and low-synchrotron peak (LSP) objects, according to the
frequency of the low-energy peak \citep{Abdo2010c}.  In most cases, BL\,Lacs
are identified as HSP, while most FSRQs are LSP. However, this
differentiation is not clear-cut since there are several transition objects.

Recently, \citet{Giommi12} studied the difference in the emission properties
between these sub-types of blazars. They found that FSRQs maintain their LSP
classification, which is the same as Fanaroff-Riley II (FRII)
radio-galaxies, according to their ionisation degree. On the other hand,
BL\,Lac objects are a mixture of two intrinsically different sources: FR\,I
objects, with weak or non-existent emission lines, and FR\,II objects whose
lines were diluted by several effects, mainly because of a strong thermal
(accretion disk) component. In this sense, the SP classification might be
driven by selection effects, while the FR\,I--FR\,II division is a
classification based on physical properties.

A detailed knowledge of the optical flux properties in blazars is thus
relevant to get a clearer picture of the complex interrelations between
emission properties at different frequencies. In particular, flux
variability studies are important to test the size and location of the
emitting region at a given frequency, as well as to give clues on the
mechanisms which originate the emission. Variability at the shortest
measurable timescales is supposed to provide information on emitting
regions at the smallest spatial scales, which can be useful to test any
relation with the emission at higher frequencies.

In this work we analyse the optical flux variability in a sample of nine
blazars. All these sources were detected at high frequencies by different
satellites in X-rays \citep[e.g.][]{Giommi12,Cusumano10}, and by \textit{Fermi} LAT
(Fermi Large Area Telescope) at GeV energies \citep{Abdo10A}. Some of them
were also detected at TeV energies by experiments like HESS and
MAGIC. According to \citet{Abdo10B}, all the blazars in our sample show a SED of the LSP
class, with the exception of PKS\,0048$-$097, which is classified as ISP.

\begin{table*}

\caption{Sample}
\label{Tab_1}

\centering
\begin{tabular}{cccccc}
\hline \hline \noalign{\smallskip}
 Object & Type & $\alpha_{\mathrm{J2000.0}}$ & $\delta_{\mathrm{J2000.0}}$ & $m_{\mathrm{NED}}$ & $z$ \\
\noalign{\smallskip} \hline \noalign{\smallskip}
 PKS\,0048$-$090    & BL-Lac      & 00:50:41.3 & $-09$:29:05 &  17.4  &  0.634\\
 PKS\,0754$+$100    & BL-Lac      & 07:57:06.6 & +09:56:35 &  14.5  &  0.266\\
 $[$HB89$]$\,0827$+$243 & Blazar (LPQ) & 08:30:52.1 & +24:11:00 &  17.2  &  0.940\\
 PKS\,0851$+$202    & BL-Lac      & 08:54:48.8 & +20:06:31 &  14.0  &  0.306\\
 PKS\,1253$-$055    & FSRQ (TeV)  & 12:56:11.1 & $-05$:47:22 &  15.2  &  0.536\\
 PKS\,1510$-$089    & QSO (HP+TeV) & 15:12:50.5 & $-09$:06:00 &  18.2  &  0.360\\
 PKS\,1749$+$096    & BL-Lac      & 17:51:32.8 & +09:39:01 &  16.8  &  0.320\\
 PKS\,2230$+$114    & FSRQ        & 22:32:36.4 & +11:43.50 &  17.3  &  1.037\\
 PKS\,2251$+$158    & FSRQ        & 22:53:57.7 & +16:08:53 &  16.1  &  0.859\\
\noalign{\smallskip} \hline
\end{tabular}

\caption*{Column 1 is the source name; column 2, the type; position is given in columns 3 and 4; visual magnitude and redshift (as found in NASA Extragalactic Database - NED) are presented in columns 5 and 6, respectively.}
\end{table*}

\section{Observations and data reduction}

A sample of nine gamma-ray blazars was photometrically followed with high temporal resolution ($<1$ hour), each object being observed along 1 to 4 (mostly) consecutive nights. The whole campaign spanned 13 months, from March 2006 to April 2007.
 The general information on the sources is shown in Table~\ref{Tab_1}. Optical images were obtained using standard $V$ and $R$ Johnson-Kron-Cousins filters with the no longer operating 1.52\,m telescope at the Estaci\'on de Observaci\'on de Calar Alto, part of the National Astronomical Observatory (OAN) of Spain. This telescope was equipped with a
1024 $\times$ 1024 pixel CCD, resulting in a field of view (FOV) of
$6'.9\times 6'.9$ (i.e., 0.40 arcsec pixel$^{-1}$ scale). Full width at half maximum (FWHM) values varied from $2.3$~arcsec to $3.7$~arcsec in some of the frames; exposure
times ranged from 100 to 840 seconds, depending on the night conditions and
the brightness of the source.

All images were de-biased and flat-fielded using the standard IRAF%
\footnote[2]{IRAF is distributed by the National Optical Astronomy
  Observatories, which are operated by the Association of Universities for
  Research in Astronomy, Inc., under cooperative agreement with the National
  Science Foundation.} reduction package. The IRAF \textsc{apphot} package
was used to extract the photometry for all the images, with aperture
diameters of $4$ arcsec for all objects with the exception of
PKS\,2230$+$114 ($3.2$ arcsec) and 0827$+$243 ($4.8$ arcsec). The size of
the aperture diameter was chosen according to the stabilisation of the
photometric growth curve. 

A comparison and a control star were chosen within each field to
build the differential light curves. These were statistically
analysed to study their variability by means of the $F$-test. For each light
curve, we estimated the $F$ parameter, defined as the ratio of the variance
of the object$-$comparison light curve ($\sigma^2_\mathrm{o-c}$) to the
variance of the control$-$comparison light curve
($\sigma^2_\mathrm{k-c}$). If $F = \sigma_\mathrm{o-c}^2 /
\sigma_\mathrm{k-c}^2 \ge F_n^{\alpha}$, the object is said to be variable,
being $F_n^{\alpha}$ a critical value. This critical value is calculated for
a set of $n=N-1$ degrees of freedom, where $N$ is the number of points in
the light-curve, while $\alpha$ is chosen to determine the desired level of
confidence. If $\alpha=0.01$, then the $F$ test has a 99\% confidence level.

In this process, a statistical weight $\Gamma$, as introduced by
\citet{Howell88}, was adopted for the $F$ test to consider effects caused
by the difference in magnitude of the source and the stars. The $\Gamma$ factor can be derived from measurable quantities from a given observation, such as sky--substracted counts from the object, sky photons, and the read--out noise.

As a result of taking this weight into account, $F=\sigma^2_\mathrm{o-c}/\left(\Gamma^2
\sigma^2_\mathrm{k-c}\right)$. The importance of this weighting to avoid spurious
variability results in differential photometry studies of blazars has been
underscored by \citet{Cellone07}.

\section{Results}

In this section we present a brief description of each object and the optical
flux variability results that we found.

\paragraph{PKS\,0048$-$097:} is a strongly variable BL-Lac object. Its
redshift is highly uncertain since its spectrum is basically flat;
\citet{Stickel93} derived a lower limit of $z \ge 0.2$. A
tentative value of $z \approx 0.63 $ was estimated \citep{RectorStocke01} and then confirmed at $z=0.635$ \citep{Landoni12}. This BL-Lac
object, as most of its kind, shows variability at almost every wavelength
range, from radio to optical bands \citep{Pica88}. It has also been detected
in X-rays \citep{Brinkmann00}. On the other hand, several works have found
variability in optical polarization, reporting a value of polarization
degree $\langle P_V \rangle=10.6\%$ \citep{Wills92}.  Through a very long
campaign (from 1979 to 2004), \citet{Kadler06} detected evidence for
long-term periodicity in radio emission, suggesting a period of $\sim 450$
days for the first half of the 80s, and a period of almost $600$ days for
late 80s and early 90s (these periods were determined with a very high
confidence level). These authors also reported a change of about
$90^{\circ}$ in the direction of the jet during its propagation from 1995 to
2002. This behaviour implies a strong structural variability, which would
deserve further studies. We used the magnitudes of the field stars published in
\citet{Villata98} to estimate the mean standard magnitude for PKS\,0048$-$097 in each filter during our observations; resulting in $15.84 \pm 0.05$ mag in $V$, $15.44 \pm 0.04$ in $R$ band the first night, and $15.89 \pm 0.05$ mag in $V$, $15.99 \pm 0.05$ in $R$ band the second night. These magnitudes are in agreement with those presented by \citet{FanLin00}, who found $V=15.86$ and $R=15.40$.  In the present work, this object showed small amplitude variability in the $R$ band.

\paragraph{PKS\,0754$+$100:} is a highly polarized and variable LSP blazar
\citep{Ghosh95,Ghosh00} of the BL-Lac type \citep{Tapia77}. Its long-term
variability at optical bands can be traced back to 1980
\citep[e.g.][]{Baumert80,Pica88,Sillanpaa91,Katajainen00}. Infrared
variations were detected by \citet{FanLin99}, as well as a radio flare by
\citet{Nieppola09}. This object shows a featureless spectrum and its host
galaxy can be resolved in the images \citep{Abraham91,Falomo96}, but it is
rather faint compared to the AGN. \citet{Nilsson03} suggest a host magnitude
of $R=18.55$\,mag and $r_\mathrm{eff}=1.3$\,arcsec.  \citet{Carangelo03}
suggest a redshift $z=0.266\pm0.001$ based on the identification of two
faint emission lines. We followed this source during the night of March
24, 2006, in which, the $F$ test results in no variability neither in the
$V$ nor in the $R$ bands. Using the comparison stars and standard magnitudes published by \citet{Fiorucci98}, the mean standard magnitudes for PKS\,0754$+$100 are $16.72 \pm 0.05$ mag in $V$ and $16.22 \pm 0.05$ mag in $R$ bands. Compared with \citet{FanLin00}, who present mean values of $V=15.40$ and $R=14.27$, our observations clearly correspond to a minimum activity state.

\paragraph{[HB89]\,0827$+$243:} is classified as a FSRQ \citep{Healey07}. It
is also a gamma-ray bright quasar, with a redshift $z=0.939$
\citep{Hewett10}. Variability on a day-long scale has
been detected at $B$, $V$, and $R$ optical bands \citep{Villata97,Raiteri98},
as well as long-term variability in the near-infrared $J$, $H$, and $K'$
bands \citep{Enya02}.  The jet emerging from the core of 0827$+$243 has been
detected both at X-ray and radio bands, showing an apparent superluminal
motion of over $20 c$ \citep{Jorstad01, Piner06}. This jet has a highly
unusual morphology bending almost $90^{\circ}$ at the X-ray band, while in
radio only the external $90^{\circ}$-bended section is visible. A
swinging-nozzle model has been proposed to explain this unusual jet
behaviour \citep{JorstadMarscher04}; this model proposes a scenario in
which the jet flow has always been straight, but the AGN smoothly changed
its direction, leaving traces of energised matter towards its former
direction, and hence the difference both in structure and in emission can be
explained. Our data show no intranight variability (both in $V$ and in $R$
bands) during each of the three nights $[$HB89$]$\,0827$+$243 was
observed. Regarding an internight time-scale, the light-curves analysis
detects significant variability.

\paragraph{PKS\,0851$+$202:} (also known as OJ\,287) is a well known AGN. Its
redshift is well-determined at $z=0.306$ \citep{Nilsson10}, and it is
usually classified as a FSRQ \citep{Healey07}. OJ\,287 shows a periodical
behaviour in its optical emission. Analysing its historical light-curve,
several studies concluded that it shows a period of $\sim 12$ years
\citep{Sillanpaa88,Sillanpaa96}, explained by a binary black hole system
\citep{Sillanpaa88,Valtaoja00,Valtonen08}. Variability has also been
detected both at radio and X-ray wavelengths
\citep{Urry96,Hughes98,Hovatta08}. In recent years a host galaxy has been
observed for this object. \citet{Nilsson03} found a diffuse host of
$R=18.9$\,mag and $r_{e}=1.0$\,arcsec, with an upper limit of $M_R > -24.0$.
The complex radio structure shown by the jet of this source was reported by
\citet{Jorstad01} and \citet{Jorstad05}. We found this source to display
strong activity, resulting variable in both filters ($R$ and $V$), in
each night and when considering two consecutive nights. With the magnitudes given by \citet{Fiorucci96} and \citet{GonzalezPerez01}, we obtained mean standard estimates of $14.93 \pm 0.01$ in $V$ and $14.47 \pm 0.01$ in $R$ the first night, and $14.91 \pm 0.01$ in $V$, $14.44 \pm 0.01$ in $R$ the second night. \citet{FanLin00} propose a periodic light curve for PKS\,0851$+$202, with a period of 12 years. Our results are in agreement with the magnitudes expected for this periodic variability and with the values observed by \citet{Bach07}.

\begin{figure*}
\centering
\includegraphics[scale=0.35]{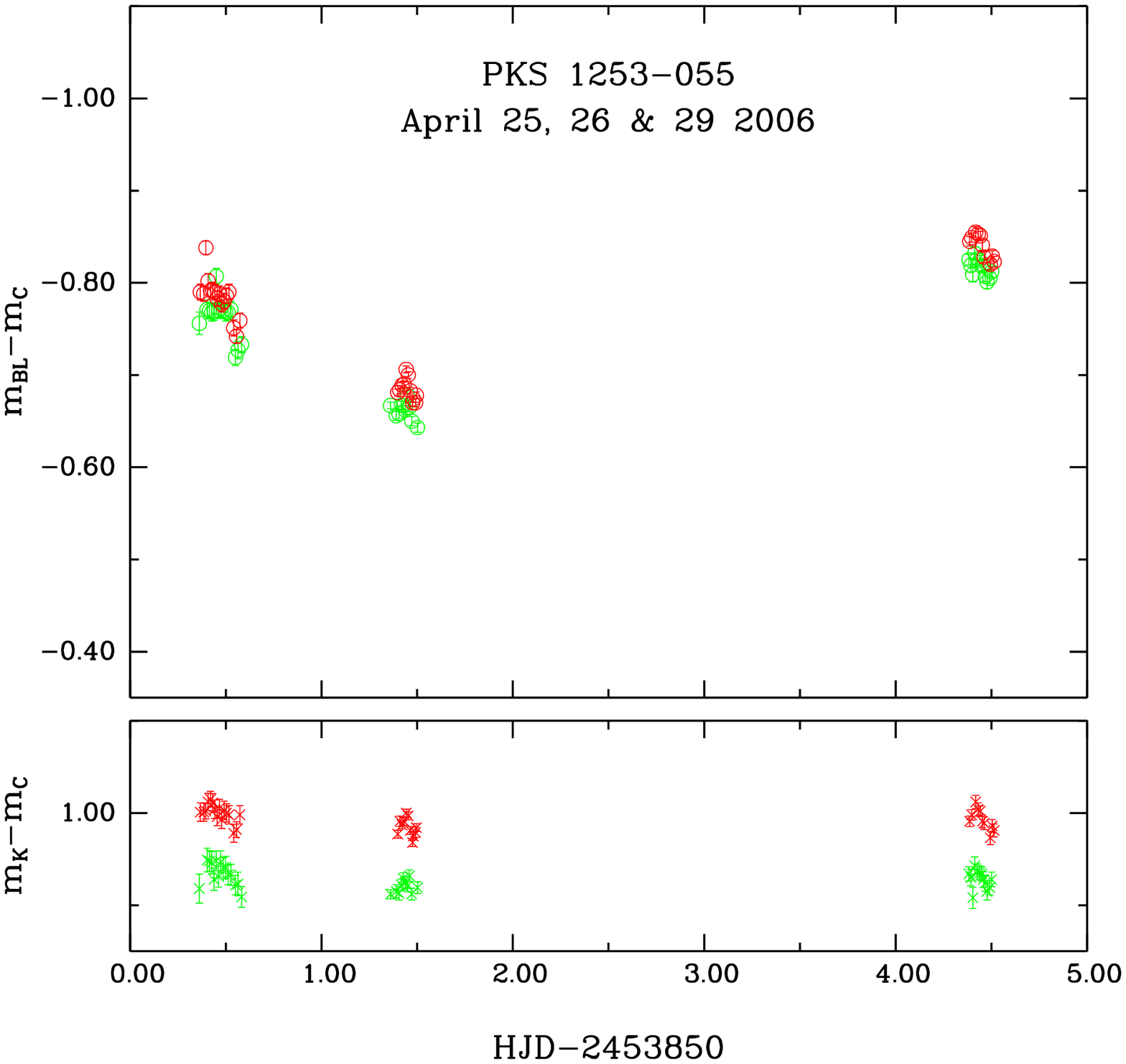}
\qquad \includegraphics[scale=0.35]{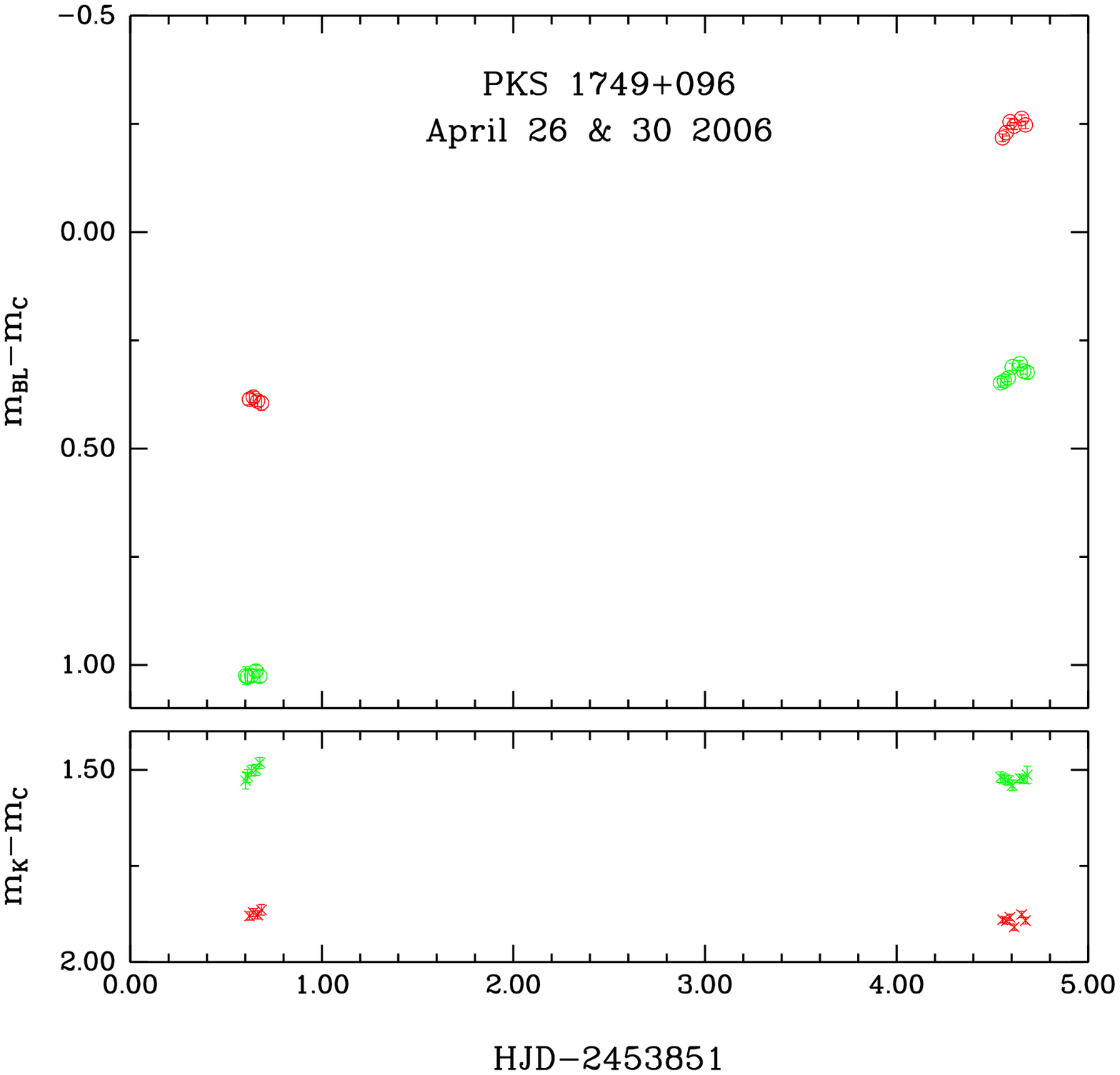} \\
\includegraphics[scale=0.35]{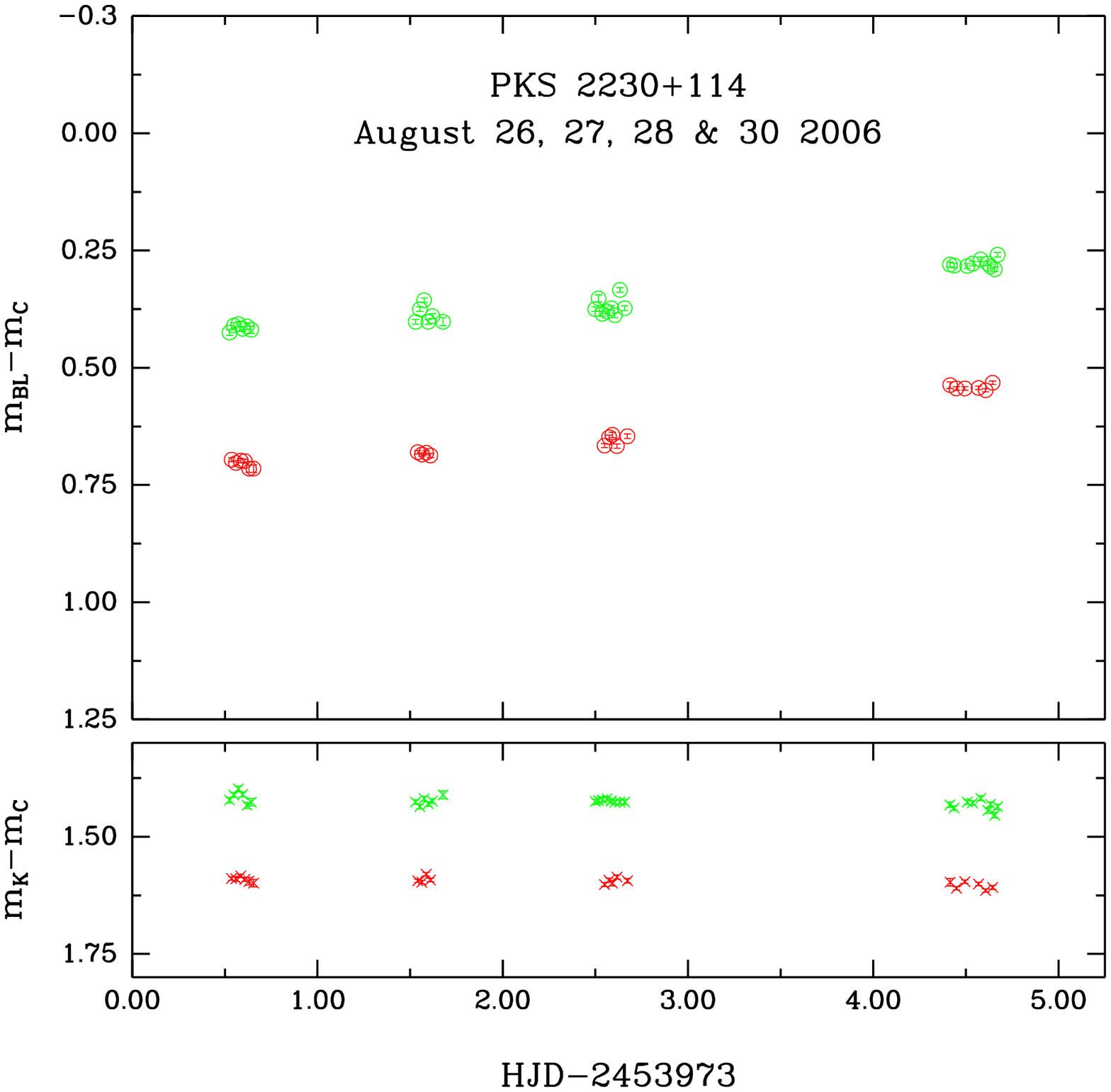}
\qquad \includegraphics[scale=0.35]{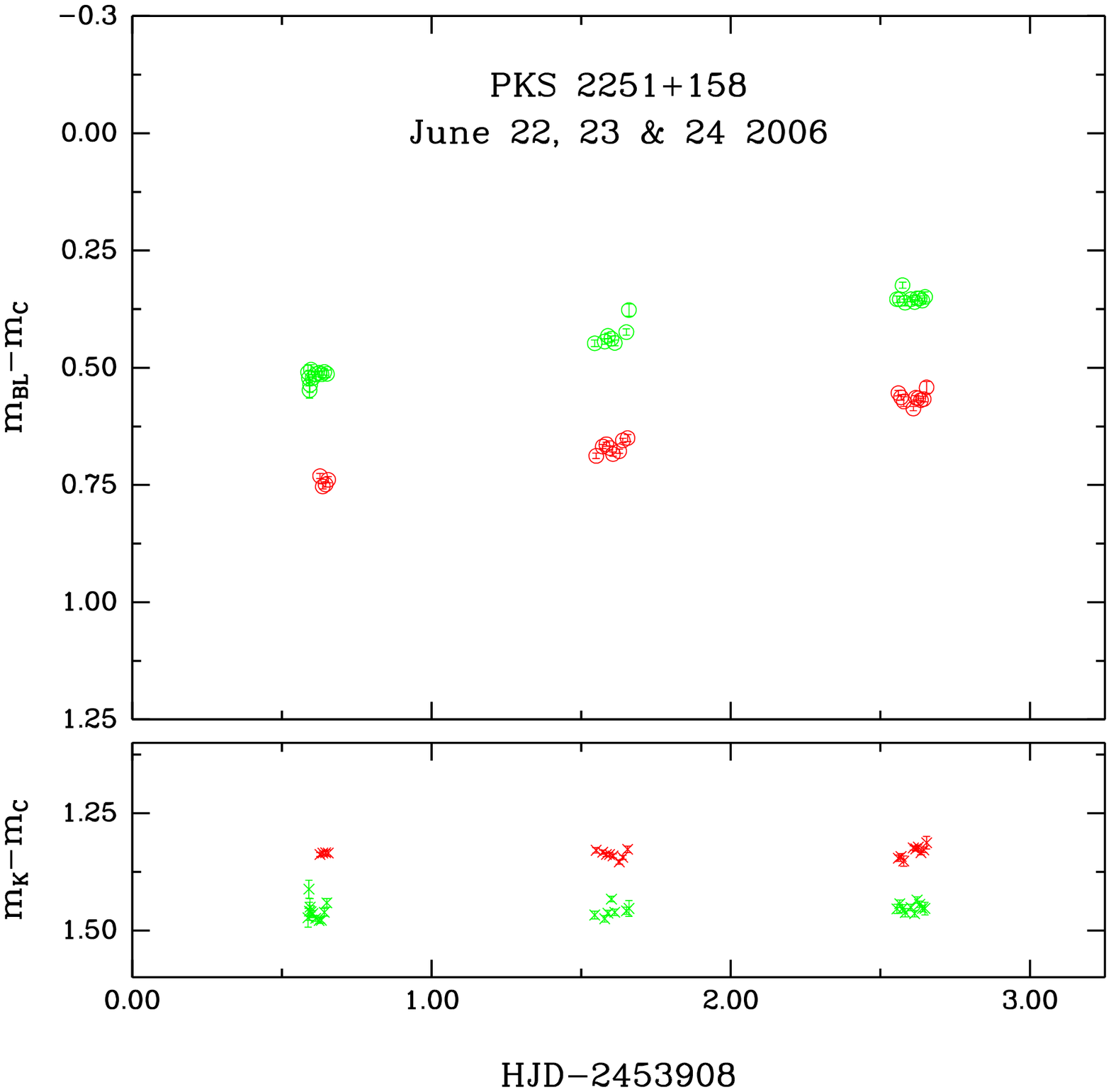} \\
\caption{Examples of differential light curves from our sample. $m_\mathrm{BL}$, $m_\mathrm{c}$,
$m_\mathrm{k}$ are the instrumental magnitudes for the blazar, the comparison
star and the control star, respectively. From top to
  bottom, left to right: PKS\,1253$-$055, PKS\,1749$+$096, PKS\,2230$+$114,
  and PKS\,2251$+$158. Data for the $V$ filter are shown in green, data for
  the $R$ filter in red. Upper panels: object $-$ comparison star
  differential light curves; lower panels: control star $-$ comparison star
  differential light curves. Note that both panels are shown with the same
  scale for a given blazar.}
\end{figure*}

\paragraph{PKS\,1253$-$055:} (also known as 3C\,279), at a redshift of $z=0.538$,
was the first blazar detected at $\gamma$-rays by the \textit{EGRET} telescope
\citep{Hartman92}, as well as the first superluminal blazar ever detected
\citep{Whitney71}. It is very well known for its variability at every
wavelength and also for its $\gamma$-ray flares \citep[see][]{Grandi96,
  Wehrle01, Bottcher07, Giuliani09}. It seems that some periods of
extreme variability could be present in this source, for example, the
optical outburst detected during 2001--2002 reported by
\citet{Kartaltepe07}.  \citet{Andruchow03} found strong microvariability in
both optical polarization degree and its position angle, and consequently
proposed a jet rotation model to explain the behaviour of the position
angle. On the other hand, \citet{Nilsson09} resolved the host galaxy,
reporting an apparent magnitude $m= 18.4\pm0.3$\,mag in the $I$ band, an
effective radius $r_\mathrm{eff}=2.7\pm1.1$\,arcsec, and a luminosity
$M_R=-23.8$\,mag.

As 3C\,279 seems to usually undergo periods of extreme variability in one
band together with extremely low variability in another band
\citep[see, e.g.][]{Collmar10}; theories that describe its SED should also
consider an anti-correlation in variability states at different wavelengths.

On February 23, 2006, this object was detected by MAGIC at very high
energies (VHE), in the TeV range \citep{Albert08}. It is interesting to note
that 3C\,279 is one of the objects with the highest well-determined redshift detected so far at VHE. Observations covering a span of four months
beginning in December 2005, reported by \citet{Bottcher07} and
\citet{Bottcher09}, show that 3C\,279 was in a high-brightness state around
the date of the $\gamma$ flare, after which its flux began to decrease at
all optical wavelengths. Our observations started less than a month after the
end of that campaign. Using the field star magnitudes published by \citet{GonzalezPerez01}, we obtained mean standard magnitudes of $V = 15.2 \pm 0.03$ and $R =14.7 \pm 0.02$\,mag $(\langle V-R
\rangle \simeq 0.5)$.

\citet{Bottcher07} report a strong flux decrease ($\Delta V/\Delta t \approx
0.15$ and $\Delta R/\Delta t \approx 0.10$\,mag per day) at the end of the
2005-2006 WEBT\footnote{The Whole Earth Blazar Telescope
  (\url{http://www.oato.inaf.it/blazars/webt/})} campaign on 3C\,279.

Our results show that the blazar continued its decreasing flux trend,
although either with a smaller slope, or the flux stabilised sometime
before the start of our observations.  We found some low amplitude activity
in both bands during the first night, but no further significant intranight
variability. However, we found strong internight variability in both bands. \\

\paragraph{PKS\,1510$-$089:} at a redshift $z=0.36\pm0.002$
\citep{Thompson90}, is classified as a highly polarized FSRQ
\citep{VeronCetty01,DAmmando09}. Its emission spans from radio wavelengths
to gamma-rays, including optical and X-ray bands
\citep{Kataoka08,Cusumano10}. This quasar has an extensive history of long and
short-term variability at optical bands \citep[see, for
example,][]{Liller75, Pica88, Villata97, Romero02, Cellone07}, with
continuous changes in brightness over long periods of time. Moreover, optical
polarization has been detected with variability both in $P$ and in position
angle $\theta$ \citep{Andruchow05}, as well as a rapid variation in
$\gamma$-rays \citep{DAmmando09}. Simultaneously with this $\gamma$-ray
flare, these authors also found intense variability at optical and radio
wavelengths. Its jet is detected primarily in radio, with superluminal
motion \citep{Homan01}, and also in X-rays \citep{Sambruna04} and at
infrared wavelengths \citep{Rantakyro98}, and is described as a rather
extended, diffuse, and bended jet \citep{Jorstad01}.  Although there was no
detectable activity during the two nights we followed this object, and no
internight variability in the $R$ band, we found clear internight
variability in the $V$ band. Again, derived from the magnitudes in \citet{GonzalezPerez01}, we found mean standard magnitudes of $16.79 \pm 0.01$ in $V$, $16.33 \pm 0.01$ in $R$ the first night, $16.73 \pm 0.02$ in $V$ and $16.26 \pm 0.01$ in $R$ the second night.

\paragraph{PKS\,1749$+$096:} is a BL-Lac object with a redshift $z=0.32$
according to \citet{Stickel88} and \citet{White88}. Its radio jet was
originally detected by \citet{Lobanov00}. This source presents a history of
radio flaring, possibly with a long-term period
\citep{Hovatta08,Nieppola09}. Flux variability was also detected at optical
\citep{Stickel93} and $\gamma$-ray frequencies \citep{Abdo10A}.  The optical
and infrared polarizations also show variability \citep{Brindle86}.  We found
small amplitude activity for this object during one of the nights in the $V$
band. However, at internight scales, this source showed the largest
amplitude variability for the whole sample, reaching a change in its
differential magnitude of $\sim 0.7$\,mag. In standard magnitudes, taking mean values and using the magnitudes given by \citet{Fiorucci98}, this detected variation resulted in a decrease in magnitude from $17.94  \pm 0.04$\,mag in $V$ and $17.10 \pm 0.05$\,mag in $R$ the first night to $17.23 \pm 0.04$ in $V$ and $16.57 \pm 0.04$ in $R$ the second night. According to the light curve given in \citet{FanLin00}, these values correspond to a minimum of activity.

\paragraph{PKS\,2230$+$114:} (also known as \,CTA\,102) is a blazar at $z=1.037$.
Multi-wavelength studies have detected correlated variability, mostly with flares from radio to X-rays, including optical and near IR bands
\citep{Bach07,OstermanMeyer09,Fromm11}. CTA\,102 also shows a jet with
superluminal motions of up to $21\,c$, derived from VLBI observations
\citep[see,][]{Jorstad01,Rantakyro03,Jorstad05}.  Its history of optical
variability can be tracked back to 1973 \citep{Pica88}, with sporadic
detections afterwards \citep[see, for example,][]{Wilkes94, Ghosh00,VeronCetty01, Romero02}.  We report here a clear internight variability in both $R$ and $V$ bands. We also detected one night (over four) with
activity, in the $V$ band.

\paragraph{PKS\,2251$+$158:} (also known as 3C\,454.3) with $z=0.860$, is one of the
most active blazars at high energies \citep{Hartman92}; during 2005 it
underwent a major flaring in almost all energy bands
\citep{Giommi06}. During 2007, a multiwavelength study was carried out by
several observatories from radio to $\gamma$-rays. In December 12, the flux
increased $\sim 1.1$\,mag in $1.5$~h, followed by a decrease of $\sim 1.2$\,
mag in $\sim 1$\,h at optical wavelengths \citep{Raiteri08}, being one of
the fastest variations ever detected in blazars. A new SED has been
determined using \textit{AGILE}/Swift data \citep{Vercellone08}. Another outburst was
detected in 2009, at optical, X-ray and gamma frequencies. This flare was
studied by \citet{Striani10}, and its polarization was studied by
\citet{Sasada12}; these authors report two distinct rotation
events. Because of its frequent outbursts and extreme variability, this
object is usually used as an example when studying the validity of the SP
classification \citep{GopalKrishna11,Ghisellini11}. We report a strong
internight variability in the $R$ and $V$ bands. Furthermore, we found
microvariability in the $R$ band within one out of three nights we followed
this blazar. Taking mean values and using the magnitudes published in \citet{GonzalezPerez01} and in \citet{Fiorucci98}, we found mean standard magnitudes of $16.37 \pm 0.09$ in $V$, $16.06 \pm 0.09$ in $R$ (first night), $16.3 \pm 0.09$ in $V$, $15.99 \pm 0.09$ in $R$ (second night), and $16.2 \pm 0.09$ in $V$, $15.88 \pm 0.09$ in $R$ (third night). These values are in agreement with those given in \citet{Raiteri07}, as with the general trend of increasing magnitudes.

\begin{table*}
\caption{Statistical results for the differential light curves.}
\label{tab:stat}
\centering

\begin{tabular}{c cc cc rr cc cc rr}
\noalign{\medskip} \hline \hline \noalign{\smallskip} UT Date & \multicolumn{2}{c}{$\sigma$~(mag)} & \multicolumn{2}{c}{Variable?} &\multicolumn{2}{c}{$F$} & \multicolumn{2}{c}{$\Gamma$} & \multicolumn{2}{c}{$N$} & \multicolumn{2}{c}{$F^t_{99\%}$}\\

~[m/d/y] &$V$ & $R$ & $V$ & $R$ & $V$~~ & $R$~~ & $V$ & $R$ & $V$ & $R$ & $V$ ~~& $R$~~ \\

\noalign{\smallskip}\hline \noalign{\smallskip}

\multicolumn{9}{l}{\rm{\textbf{PKS\,0048-090}}}\\
09/26/06 & 0.050 &0.029 & NV & NV & 2.36 & 1.49 & 0.74 & 0.89 & 18 & 19 & 3.24 & 3.13 \\
09/27/06 & 0.030 &0.046 & NV & NV & 2.42 & 1.29 & 0.73 & 0.88 & 27 & 31 & 2.55 & 2.39 \\
[2pt] All nights &0.040 & 0.040 & NV & V & 1.91 & 2.13 & 0.73 & 0.89 & 43 & 49 & 2.08 & 1.98 \\

\noalign{\smallskip} \hline \noalign{\smallskip}

\multicolumn{9}{l}{\rm{\textbf{PKS\,0754+100}}}\\
03/24/07 & 0.024 &0.024 & NV & NV & 4.35 & 1.89 & 1.04 & 0.99 & 11 & 10 & 4.85 & 5.35 \\

\noalign{\smallskip}\hline \noalign{\smallskip}

\multicolumn{9}{l}{\rm{\textbf{$[$HB89$]$\,0827+243}}}\\
03/20/07 &0.022 & 0.010 & NV & NV & 1.12  & 5.47 & 1.18 & 1.21 & 05 & 04 & 15.97 & 29.44 \\
03/22/07 &0.013 & 0.009 & NV & NV & 1.24 & 1.32 & 1.12 & 1.14 & 12 & 16 & 4.46 & 3.52 \\
03/23/07 &0.016 & 0.013 & NV & NV & 1.45 & 3.16 & 1.08 & 1.12 & 12 & 14 & 4.46 & 3.90  \\
[2pt] All nights & 0.016 & 0.013 & V & V & 6.07 & 6.86 & 1.10 & 1.13 & 29 & 33 & 2.46 & 2.32\\

\noalign{\smallskip} \hline \noalign{\smallskip}

\multicolumn{9}{l}{\rm{\textbf{PKS\,0851+202}}}\\
03/28/06 & 0.006 &0.007 & V & V & 4.77 & 5.80 & 0.98 & 0.93 & 31 & 30 & 2.39 & 2.42\\
03/29/06 & 0.007 &0.007 & NV & V & 2.73 & 5.12 & 0.98 & 0.91 & 13 & 12 & 4.16 & 4.46\\
[2pt] All nights &0.007 & 0.007 & V & V & 6.27 & 8.58 & 0.99 & 0.98 & 44 & 42 & 2.06 & 2.09\\

\noalign{\smallskip} \hline \noalign{\smallskip}

\multicolumn{9}{l}{\rm{\textbf{PKS\,1253-055}}}\\
04/25/06 & 0.009 & 0.008 & V & V & 6.59 & 7.65 & 0.67 & 0.81 & 15 & 16 & 3.70 & 3.52 \\
04/26/06 & 0.010 & 0.008& NV & NV & 4.18 & 2.14 & 0.68 & 0.82 & 10 & 11 & 5.35 & 4.85 \\
04/29/06 & 0.007 & 0.009 & NV& NV & 2.44 & 2.03 & 0.66 & 0.80 & 12 & 10 & 4.46 & 5.35 \\
[2pt] All nights & 0.008 & 0.010 &V & V & 63.05 & 42.49 & 0.67 & 0.81 & 37 & 37 & 2.21 & 2.21\\

 \noalign{\smallskip} \hline\noalign{\smallskip}

\multicolumn{9}{l}{\rm{\textbf{PKS\,1510-089}}}\\
06/03/06 & 0.009 &0.025 & NV & NV & 3.27 & 1.04 & 1.02 & 1.03 & 08 & 06 & 7.00 & 10.97\\
06/04/06 & 0.007 &0.018 & NV & NV & 6.22 & 1.23 & 0.97 & 0.98 & 08 & 05 & 7.00 & 15.97 \\
[2pt] All nights &0.009 & 0.021 & V & NV & 14.95 & 3.18 & 0.97 & 1.02 & 13 & 10 & 4.16 & 5.35\\

\noalign{\smallskip} \hline \noalign{\smallskip}

\multicolumn{9}{l}{\rm{\textbf{PKS\,1749+096}}}\\
04/26/06 & 0.018 &0.007 & NV & NV & 11.11 & 2.45 & 1.00 & 0.99 & 05 & 04 & 15.97 & 29.44\\
04/30/06 & 0.006 &0.007 & V & NV & 8.56 & 4.55 & 0.68 & 0.67 & 07 & 06 & 8.47 & 10.97\\
[2pt] All nights & 0.011& 0.009 & V & V & 1902.39 & 1411.39 & 0.73 & 0.69 & 12 & 10 & 4.46 & 5.35\\

\noalign{\smallskip}\hline \noalign{\smallskip}

\multicolumn{9}{l}{\rm{\textbf{PKS\,2230+114}}}\\
08/26/06 & 0.013 &0.007 & NV & NV &  3.70 & 2.28 & 1.00 & 1.08 & 06 & 06 & 10.97 & 10.97\\
08/27/06 & 0.008 &0.008 & NV & NV &  4.95 & 4.31 & 0.98 & 1.07 & 06 & 04 & 10.97 & 29.44\\
08/28/06 & 0.003 &0.006 & V & NV &  46.01 & 3.23 & 0.97 & 1.04 & 08 & 05 & 7.00 & 16.00\\
08/30/06 & 0.010 & 0.007& NV & NV &  1.63 & 1.97 & 0.91 & 0.95 & 09 & 06 & 6.03 & 10.97\\
[2pt] All nights & 0.011 &0.009 & V & V & 29.50 & 62.29 & 0.95 & 1.02 & 29 & 21 & 2.46 & 2.94 \\

 \noalign{\smallskip}\hline \noalign{\smallskip}

\multicolumn{9}{l}{\rm{\textbf{PKS\,2251+158}}}\\
06/22/06 & 0.015 &0.002 & NV & V & 3.57 & 37.55 & 0.79 & 0.94 & 07 & 04 & 8.47 & 29.44 \\
06/23/06 & 0.019 & 0.008& NV & NV & 2.01 & 3.03 & 0.74 & 0.89 & 09 & 08 & 6.03 & 7.00 \\
06/24/06 & 0.006 & 0.011 & NV& NV & 2.81 & 1.54 & 0.71 & 0.83 & 10 & 09 & 5.35 & 6.03\\
[2pt] All nights & 0.011 & 0.009 &V & V &  44.68 & 65.45 & 0.76 & 0.89 & 26 & 21 & 2.60 & 2.94 \\

 \noalign{\smallskip} \hline\noalign{\smallskip}
\end{tabular}

\end{table*}

The variability results are summarised in Table~\ref{tab:stat}\footnote{Photometry results summarised in Table 2 are only available in electronic form
at the CDS via anonymous ftp to cdsarc.u-strasbg.fr (130.79.128.5) or via http://cdsweb.u-strasbg.fr/cgi-bin/qcat?J/A+A/}. Column 1
gives the object name with the date, while the following columns give,
respectively for $V$ and $R$: the observational error, $\sigma$, obtained
from the standard deviation of the control$-$comparison differential
light-curve for each filter; the variability results; the confidence
parameter ($F$); the gamma corrective factor ($\Gamma$); the number of points
in the light curves; and the corresponding critical value, $F^t$ for $n=N-1$
degrees of freedom, assuming a $99\%$ confidence level. As can be seen,
from a total of 60 light curves (each night and all the nights together for
each object), $35\%$ resulted variable at the $99\%$ confidence level. If we
relax the confidence level to a $95\%$, this percentage increases up to
$47\%$. In Fig. 1 we show the differential light curves for
PKS\,1253$-$055, PKS\,1749$+$096, PKS\,2230$+$114, and PKS\,2251$+$158, as an
example for the whole campaign; $m_\mathrm{BL}$, $m_\mathrm{c}$,
$m_\mathrm{k}$ are the instrumental magnitudes for the blazar, the comparison
star and the control star, respectively.  Observations in the $R$ band are
represented in red, while observations in the $V$ band are shown in green.

As a check, we re-calculated the variability state of each light curve after
exchanging the stars, using the comparison as a control star and
vice versa. The results did not change, except for four light curves which
changed their variability state.
This is an indication that in most cases we were able to choose well
behaved stars as comparison and control. Another indirect evidence of
this is the fact that $\Gamma \approx 1$ in most cases.
In general, $\sigma$ values were below $0.02$\,mag with the exception of
PKS\,0048$-$090, with $\sigma=0.04$.

\section{Discussion}

Internight variability was detected in most of our sample. In particular, results for PKS\,1749$+$096, PKS\,1253$-$055, PKS\,2230$+$114,
and PKS\,2251$+$158 show the largest amplitudes. The most notable variation
was detected in PKS\,1749$+$096, resulting in an amplitude $\Delta m \sim
0.7$\,mag along almost five days.

According to the $F$ test results, the largest short-scale variations were detected for
PKS\,2230$+$114 and PKS\,2251$+$158, one in each band, with an amplitude of
0.05\,mag in the first case and of 0.02\,mag in the second. However, given that the other band shows no signs of variability and that there are few data points, we suggest taking the $F$-test results with caution (see Zibecchi et al. in preparation, for a critical evaluation of the F--test and other statistics as variability indicators).

Different models have been proposed to explain (micro)variability in
blazars. In particular, we want to test whether our variability results can be explained by a swinging jet scenario \citep[e.g.][and references
therein]{GopalKrishnaWiita92,Romero95,Bachev12}. This means that variability is a consequence of slight
deviations of the direction of motion of the emitting particle populations along the relativistic jet, which is close to the
line of sight, thus leading to a change in the associated Doppler factor. This, in turn, produces a significant change in the observed magnitude.
 This is a consequence of relativistic effects in the jet plasma playing an
important role by magnifying any perturbation. Since PKS\,1749$+$096
presented the strongest variation measured during our campaign, we show and discuss the application of this model to our results for this specific object. We find a similar
result for the remaining objects in our sample.

Following \citet{Nesci02} the expected temporal change in the observed
magnitude of the blazar is related to the temporal change in the jet viewing
angle $\mathrm{d} \theta/\mathrm{d} t_\mathrm{jet}$ by the expression
\begin{equation}
 \mathrm{d} m/\mathrm{d} t_\mathrm{obs} = 1.086\,(3+\alpha)\,\beta\, \gamma\,
 \delta^2 \sin(\theta) \, (1+z)^{-1} \,(\mathrm{d} \theta/\mathrm{d}
 t_\mathrm{jet}) ,
\label{eq:nesci}
\end{equation}
where $\alpha$ is the spectral index ($F_\nu \propto \nu^{-\alpha}$), $\beta$ the bulk plasma velocity in
the jet in terms of the speed of light, $\gamma$ is the Lorentz factor,
$\delta$ is the Doppler factor, and we have included the redshift correction
factor. We note that the time interval on the left hand corresponds to the
observer's frame, while that on the right hand is in the jet's reference
frame.

To calculate $\mathrm{d}\theta/\mathrm{d}t_\mathrm{jet}$, we used the
 parameters reported by \citet{Hovatta09}: a Lorentz factor
$\gamma=7.5$, a Doppler factor $\delta=12.0$, and a jet angle to
the line of sight $\theta = 3.8^\circ$, while we adopted the
optical spectral index $\alpha_{VR} = 2.85$ from Ojha et al. (2009).

From the data here reported, we estimated a mean value of $\mathrm{d}
m/\mathrm{d} t_\mathrm{obs} = -0.16$ mag per day. Calculating the
(intrinsic) velocity, $\beta = \sqrt{1 - 1/\gamma^2} = 0.991$, we then use
Eq.\,\ref{eq:nesci} to obtain $\mathrm{d} \theta/\mathrm{d} t_\mathrm{jet} = -4.69 \times 10^{-4}$ radians per day, i.e., $\mathrm{d} \theta/\mathrm{d}
t_\mathrm{jet} = -1.6$ arcmin per day.

This procedure was also applied to another blazar not included in this
campaign, AO\,0235$+$164 ($z=0.94$). This object showed a very strong
hour-scale variability reported by \citet{Romero00a} and
\citet{Cellone07}. In this case, we used the observational data from the
night with the highest amplitude
($\mathrm{d}m/\mathrm{d}t_\mathrm{obs}=2.67$ mag per day), and another night
with a lower amplitude ($\mathrm{d}m/\mathrm{d}t_\mathrm{obs} = 1.96$ mag
per day). The relativistic parameters for this blazar are: $\gamma=12.1$, $
\delta=24.0 $, $\theta=0.4^{\circ}$ \citep{Hovatta09}, and
$\alpha_{VR}=2.15$, from \citet{Cellone07}. The results were
$\mathrm{d}\theta/\mathrm{d}t_\mathrm{jet}=1.91\times 10^{-2}$ rad per day,
i.e., $\mathrm{d}\theta/\mathrm{d}t_\mathrm{jet}=65.7$ arcmin per day in
the first, highly variable night chosen, and
$\mathrm{d}\theta/\mathrm{d}t_\mathrm{jet} = 1.40\times 10^{-2}$ rad per
day, i.e., $\mathrm{d}\theta/\mathrm{d}t_\mathrm{jet}=48.2$ arcmin per day,
in the second case.

However, we note that given a near zero value of $\theta$, such as that
reported by Hovatta et al. (2009), the relevant equations rapidly tend to a
strongly non-linear behaviour. Thus, we also used the set of parameters
inferred from a more realistic SED modelling that was given by Ackermann et
al. (2012): $\gamma = \delta = 20$, $\theta = 2.3^\circ$. These resulted in
$\mathrm{d}\theta/\mathrm{d}t_\mathrm{jet} = 2.89 \times 10^{-3}$ radians
per day, i.e., $\mathrm{d}\theta/\mathrm{d}t_\mathrm{jet} = 9.9$ arcmin per
day in the first night, and $\mathrm{d}\theta/\mathrm{d}t_\mathrm{jet} =
2.12 \times 10^{-3}$ radians per day, i.e.,
$\mathrm{d}\theta/\mathrm{d}t_\mathrm{jet} = 7.3$ arcmin per day in the
second night.

From these results, it seems that it is possible to understand the origin for the internight variability detected in PKS\,1749+096 as a slight change
in the direction of the jet-emitting region with respect to the line of
sight.  On the other hand, for variations as high as those detected in
AO\,0235+164, faster changes of the viewing angle are required (from a few
arcmin up to around a degree, depending on the adopted parameters) to
explain the variability reported as that expected as being only due to a swinging jet.

So, a considerable part of the short-term variations usually detected in blazars could
be produced by jet wiggles, mainly those sufficiently low to be associated
with a wiggle of a few arcseconds or even arcminutes. However, it is
more difficult to explain degree-scale changes in the angle of the emitting
region in less than a day, as this model would imply for AO\,0235+164,
especially if a small value of the viewing angle is adopted. On the other
hand, AO\,0235+164 did present strong changes in its colour index \citep{Romero00a}, thus further disfavouring the wiggling-jet scenario in its case.

To evaluate the rates of change in the viewing angle, we have so far
considered time intervals in the jet reference-frame. Thus, our conclusions
should be valid for small variations in the direction of the emitting
region, propagating down the relativistic jet \citep[e.g.][]{Andruchow03}. These changes can be the effect of a twisted helical magnetic field component in the inner jet. A helical field is expected beyond the Alfv\'en radius in most models of jet launching \citep[e.g.][]{Spruit10}. If we were dealing, instead, with large-scale fluctuations of the angle subtended by the jet itself (such as in a precessing-jet scenario), we
should refer the timescales to the galaxy's rest frame, thus dropping one
$\delta$ factor from Eq.\,\ref{eq:nesci}. While, in this frame, rates of
change in the viewing angle for AO\,0235+164 become implausibly high, for
PKS\,1749+096 we obtain $\mathrm{d}\theta/\mathrm{d}t_\mathrm{gal} = -19.3$
arcmin per hour, ruling out any large-scale phenomenon, but still probably
allowing for small bendings at a sub-parsec scale \citep{GopalKrishnaWiita92}.

\section{Conclusions}

We presented optical differential photometry data for a sample of nine blazars. Only four of them show statistically significant variability: PKS\,1253$-$055, PKS\,1749$+$096, PKS\,2230$+$114 and PKS\,2251$+$158.  In the particular case of PKS\,1253$-$055, we found that it was at a flux state consistent with the brightness decreasing trend reported by \citet{Bottcher09}. Moreover, lower than expected minimum brightness states were detected for PKS\,0754$+$100 and PKS\,1749$+$096.

On the other hand, the variations detected in PKS\,1749$+$096 are quite remarkable: $\sim 0.7$ mag in four days, resulting in a very high variability parameter in both bands ($F_V=1902.39$ and $F_R=1411.39$).

To explain these variability results, we applied the geometric model as in \citet{Nesci02}, which assumes that the variation in brightness responds to a slight change in the direction of the emitting region with respect to the line of sight. This change in direction, though small, can produce a noticeable change in brightness owing to a relativistic Doppler effect. Using the data given by \citet{Hovatta09}, we found this model can explain our results with jet wiggles of just $\sim 10$ arcmin per day.

\begin{acknowledgements}

E. J. M. would like to thank FCAGLP and UNLP for offering the opportunity to study this career at no cost, as well as for the resources given to accomplish this work. S. A. C. and I. A. thank ANPCyT for funding (PICT2008-0627). The whole team would also like to thank Dr. Nicola Masetti for his suggestions, as well as the anonymous referee for the clear and helpful advice given.
This work was supported by the Consejer\'{\i}a de Econom\'{\i}a, Innovaci\'on, Ciencia y Empleo of Junta de Andaluc\'{\i}a 
under excellence grant FQM-1343 and research group FQM-322, as well as FEDER funds. G. E. R. is supported by Grant AYA2013-47447-C3-1-P (Spain).
\end{acknowledgements}


\end{document}